\documentclass[a4paper,11pt]{article}
\usepackage{jinstpub} 
\usepackage{lineno}
\usepackage{scrextend}

\usepackage[lofdepth=1]{subfig}
\usepackage{float}
\usepackage{gensymb}
\usepackage{upgreek}

\title{\boldmath An extreme thermal cycling reliability test of ATLAS ITk Strips barrel modules}

\author[b,*]{A. Tishelman-Charny, \note[*]{Corresponding author.}}
\author[f]{A. Affolder,}
\author[a]{F. Capocasa,}
\author[a]{E. Duden,}
\author[f]{V. Fadeyev,}
\author[f]{M. Gignac,}
\author[e]{C. Helling,}
\author[d]{H. Herde,}
\author[f]{J. Johnson,}
\author[b]{D. Lynn,}
\author[c]{M. Morii,}
\author[k]{A. Mitra,}
\author[g,h]{L. Poley,}
\author[a]{G. Sciolla,}
\author[b]{S. Stucci,}
\author[i]{P. Sharma,}
\author[b]{G. Van Nieuwenhuizen,}
\author[d]{E. Wallin,}
\author[f]{A. Wang,}
\author[j]{and S. Wonsak}

\affiliation[a]{Department of Physics, Brandeis University, Waltham MA}
\affiliation[b]{Brookhaven National Laboratory (BNL), Upton, NY 11973 (NY), U.S.A}
\affiliation[c]{Laboratory for Particle Physics and Cosmology, Harvard University, Cambridge MA}
\affiliation[d]{Lund University, Department of Physics, Box 118, 221 00 Lund, Sweden}
\affiliation[e]{Department of Physics, University of British Columbia, Vancouver BC, Canada}
\affiliation[f]{Santa Cruz Institute for Particle Physics (SCIPP), University of California, Santa Cruz, CA 95064, U.S.A.}
\affiliation[g]{Department of Physics, Simon Fraser University, 8888 University Drive, Burnaby, B.C. V5A 1S6, Canada}
\affiliation[h]{TRIUMF, 4004 Wesbrook Mall, Vancouver, B.C. V6T 2A3, Canada}
\affiliation[i]{University of Iowa, Iowa City IA}
\affiliation[j]{Oliver Lodge Laboratory, University of Liverpool, Liverpool}
\affiliation[k]{University of Warwick, UK - Coventry CV4 7AL}

\emailAdd{abraham.tishelman.charny@cern.ch}

\abstract{At the end of Run 3 of the Large Hadron Collider (LHC), the accelerator complex will be upgraded to the High-Luminosity LHC (HL-LHC) in order to increase the total amount of data provided to its experiments. To cope with the increased rates of data, radiation, and pileup, the ATLAS detector will undergo a substantial upgrade, including a replacement of the Inner Detector with a future Inner Tracker, called the ITk. The ITk will be composed of pixel and strip sub-detectors, where the strips portion will be composed of 17,888 silicon strip detector modules. During the HL-LHC running period, the ITk will be cooled and warmed a number of times from about ${-35}^\circ$C to room temperature as part of the operational cycle, including warm-ups during yearly shutdowns. To ensure ITk Strips modules are functional after these expected temperature changes, and to ensure modules are mechanically robust, each module must undergo ten thermal cycles and pass a set of electrical and mechanical criteria before it is placed on a local support structure. This paper describes the thermal cycling Quality Control (QC) procedure, and results from the barrel pre-production phase (about 5\% of the production volume). Additionally, in order to assess the headroom of the nominal QC procedure of 10 cycles and to ensure modules don't begin failing soon after, four representative ITk Strips barrel modules were thermally cycled 100 times - this study is also described.}

\keywords{Particle tracking detectors, Si microstrip and pad detectors}

\begin{document}
\maketitle
\flushbottom

\section{Introduction}
\label{sec:introduction}

\begin{figure}[htbp]
\centering
\includegraphics[width=\textwidth]{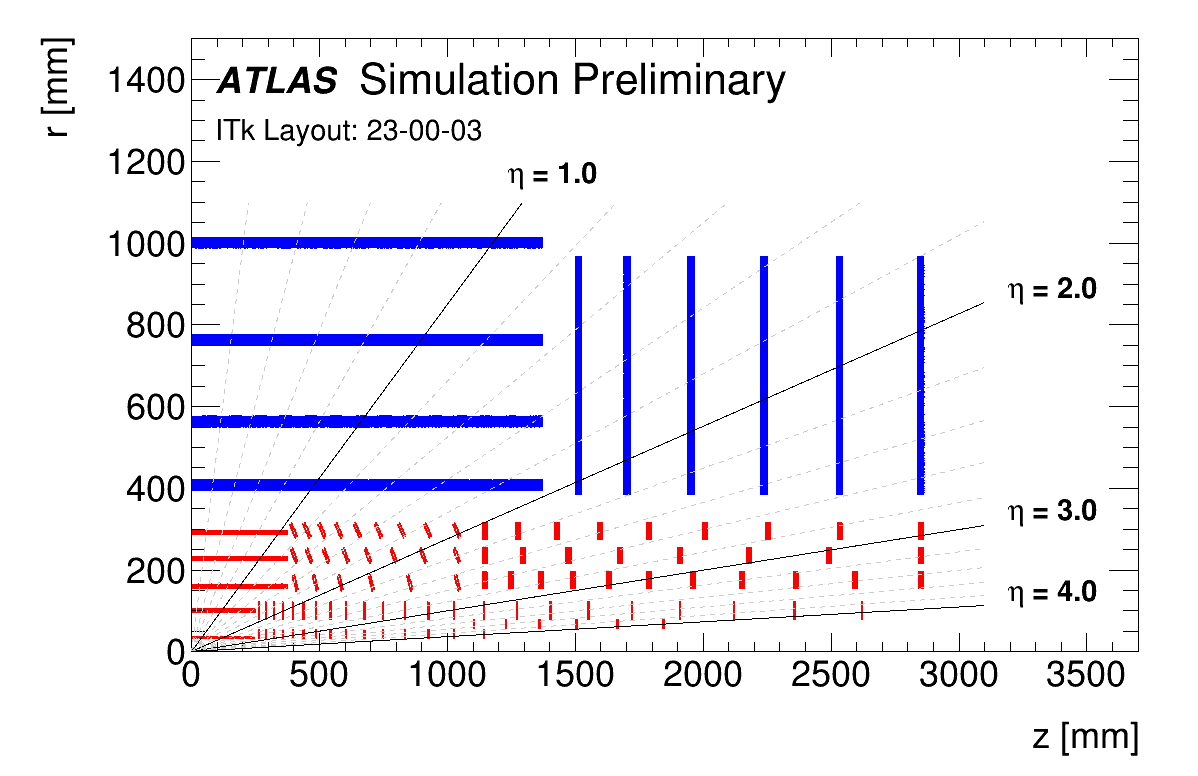}
\caption{ 
The ITk layout~\cite{ITk_Layout}.
\label{fig:ITk_layout}}
\end{figure}

During Long Shutdown 3 (LS3) of the Large Hadron Collider (LHC) running schedule, the accelerator complex will be upgraded in order to increase the instantaneous luminosity of its particle beams. This upgraded phase of the LHC, the HL-LHC (High-Luminosity LHC), is estimated to deliver 3000-4000 fb$^{-1}$ of proton-proton collisions to the LHC experiments~\cite{HLLHC_TDR}. An increase in instantaneous and integrated luminosity of this magnitude requires the LHC experiments to be upgraded to cope with higher rates of data and long term effects from radiation damage, in addition to an increased number of simultaneous interactions known as pileup. As a part of the ATLAS detector's upgrade for the HL-LHC, the Inner Detector (ATLAS tracker) will be entirely replaced by a fully silicon tracker composed of pixel and strip subdetectors, called the Inner Tracker (ITk), partitioned into a barrel and end-caps with different geometries, as shown in figure~\ref{fig:ITk_layout}.

During the full operational period of the upgraded ATLAS detector, estimated to be about ten years, the ITk will undergo a number of temperature changes as it is cooled for operations, warmed for maintenance, and experiences expected power cuts. This includes at least one expected warm-up per year due to end of the year shutdowns\footnote{However, once sufficient radiation damage has accumulated, the tracker will be kept cool, as annealing affects the impact of radiation damage~\cite{Orr:2023eb}.}. It is imperative that the detector components are not critically damaged from these temperature changes, and that all assembled modules are mechanically robust, to continue taking high quality data throughout the lifetime of the ITk.

The building block of the ITk Strips detector is a module \cite{ABC130}. Modules are composed of a silicon strip sensor, with a custom printed circuit board called a ``hybrid" hosting custom application specific integrated circuits (ASICs), and a powerboard directly glued on top. To ensure modules still operate properly after repeated temperature changes, each assembled module must pass a thermal cycling Quality Control (QC) test. This entails varying the temperature of the module from -35$\degree$C (the planned cold operating temperature during operations once sufficient radiation damage has accumulated), to +20$\degree$C, with a test suite performed at each cold and warm point, defined in section~\ref{sec:TC}. After ten thermal cycles, a module must pass a set of electrical and mechanical quality criteria before being placed on a local support structure.

In an effort to assess the headroom of the ten thermal cycle requirement of the thermal cycling QC, ensuring that modules do not begin failing soon after, four representative barrel modules assembled during the pre-production phase were thermally cycled 100 times, with electrical properties monitored throughout the testing period. This requirement of 100 cycles is also motivated by power cuts, which are expected to occur during operations. From previous experience, power cuts occur about 3-5 times per year, leading to an additional 30-50 expected cycles over a 10 year operation period. Therefore, it must be demonstrated that ITk strip modules can withstand order of 100 cycles to account for the most extreme scenario. 

This paper is structured as follows: section~\ref{sec:setup} describes an example experimental setup used for thermal cycling modules. Section~\ref{sec:TC} describes the thermal cycling QC procedure. Section~\ref{sec:Extreme_Thermal_Cycling} describes a thermal cycling reliability test where four modules were thermally cycled 100 times, and section~\ref{sec:conclusions} presents the conclusions of the paper.

\section{Experimental setup}
\label{sec:setup}

\begin{figure}[htbp]%
    \setcounter{subfigure}{0}
    \centering
    \subfloat[An example barrel module diagram \cite{TDR}.\label{fig:module_diagram}]{\includegraphics[height=0.25\textheight]{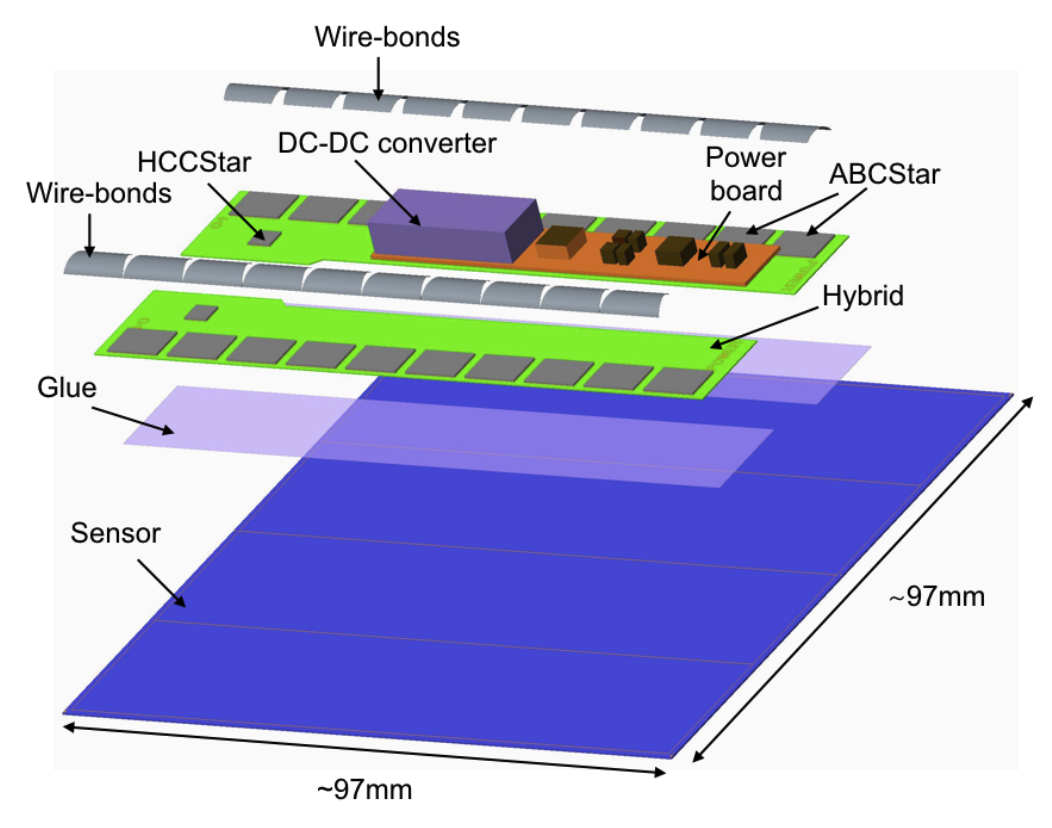}}%
    \hfill
    \subfloat[A photo of a Long Strip module~\cite{ModuleImage}, with streams 0 and 1 labelled. Note that strips are aligned in the vertical direction. \label{fig:module_image}]{\includegraphics[height=0.25\textheight]{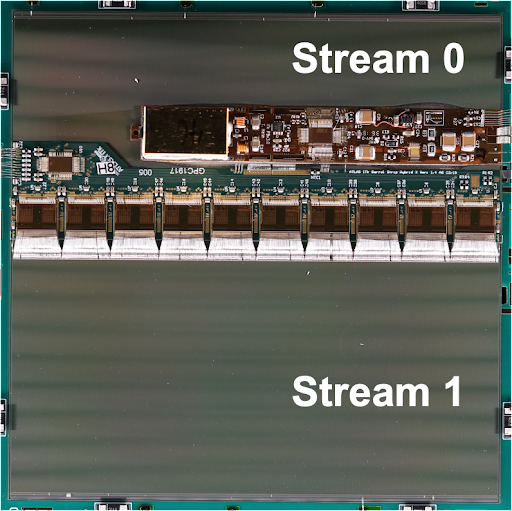}}%
    \caption{Diagram and picture of a module. }%
\end{figure}

ATLAS ITk Strips modules are built from a silicon sensor as the active detection material, with a hybrid glued directly on top hosting ASICs for readout, alongside a powerboard to power module components. The ITk Strips endcaps will be built from 6 types of modules with different geometries in order to satisfy the radial geometry requirement, while the barrel will be composed of Short Strip (SS) and Long Strip (LS) modules, whose names correspond to the modules' relative strip sizes. A diagram portraying an example barrel module is shown in figure~\ref{fig:module_diagram}. A photo of a Long Strip module, defined as a sensor with two ``streams" of strips, a single powerboard and hybrid with 10 ASICs, is shown in figure~\ref{fig:module_image}. Note that strips under (away from) the powerboard are defined as stream 0 (1).

\begin{figure}[htbp]
\centering
\includegraphics[width=\textwidth]{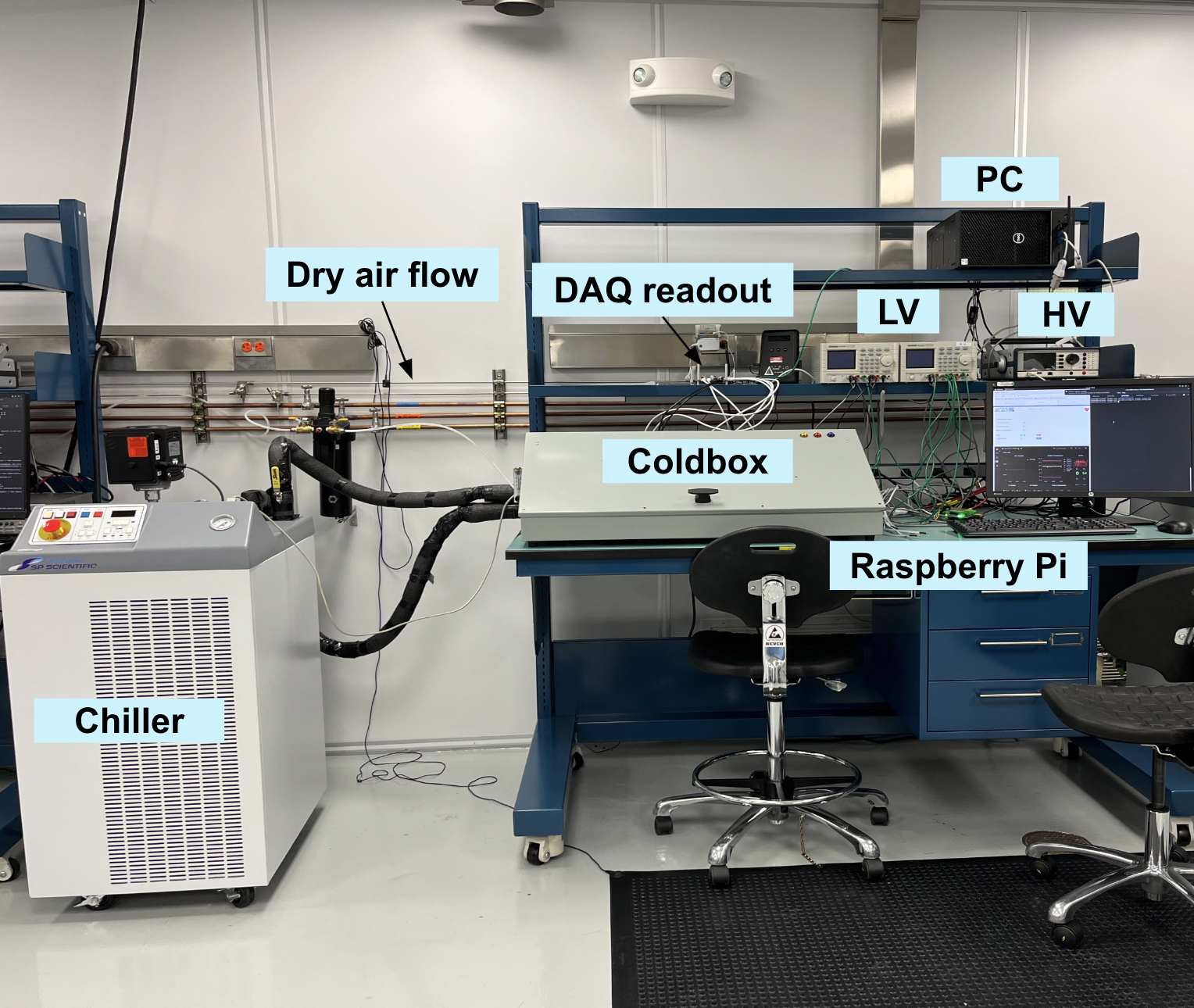}
\caption{ 
An example coldjig setup, comprised of a coldbox housing modules, and its associated hardware to control and monitor module environment and DAQ.
\label{fig:ColdjigSetup}}
\end{figure}

Thermally cycling modules requires an experimental setup which controls and monitors both environmental and data acquisition (DAQ) features. This is accomplished in the ITk Strips project with a coldjig, defined as a setup including a coldbox and associated hardware, an example of which is shown in figure~\ref{fig:ColdjigSetup}. The following are pictured: A coldbox, containing up to four modules screwed into metal plates called ``chucks". The temperatures of the chucks, and thereby the modules, is controlled with a set of cooling pipes between the chiller and the chucks. The coldbox is constantly flushed with dry air flowing from external pipes in order to minimize humidity, to prevent any ice or moisture from condensing on the modules during testing. A DAQ readout, consisting of a Nexys field programmable gate array~\cite{NEXYS} for fast processing of module data, and a custom interface board (FMC-0514-DP~\cite{FMC}) between the Nexys and modules, is used to perform various DAQ tests to measure module noise and gain at the strip level~\cite{ABC130}. Low voltage (LV) power supplies are connected to the coldjig to provide 11V to module electronics. A High voltage (HV) power supply is connected to provide -350V bias voltage for sensor depletion during electrical tests and to take I-V measurements up to -550V. A PC is present to run DAQ software, and a Raspberry Pi~\cite{RaspPi} is used for monitoring and control of the environmental elements of the setup including temperature, humidity, and dew point.

\begin{figure}[htbp]
\centering
\includegraphics[width=\textwidth]{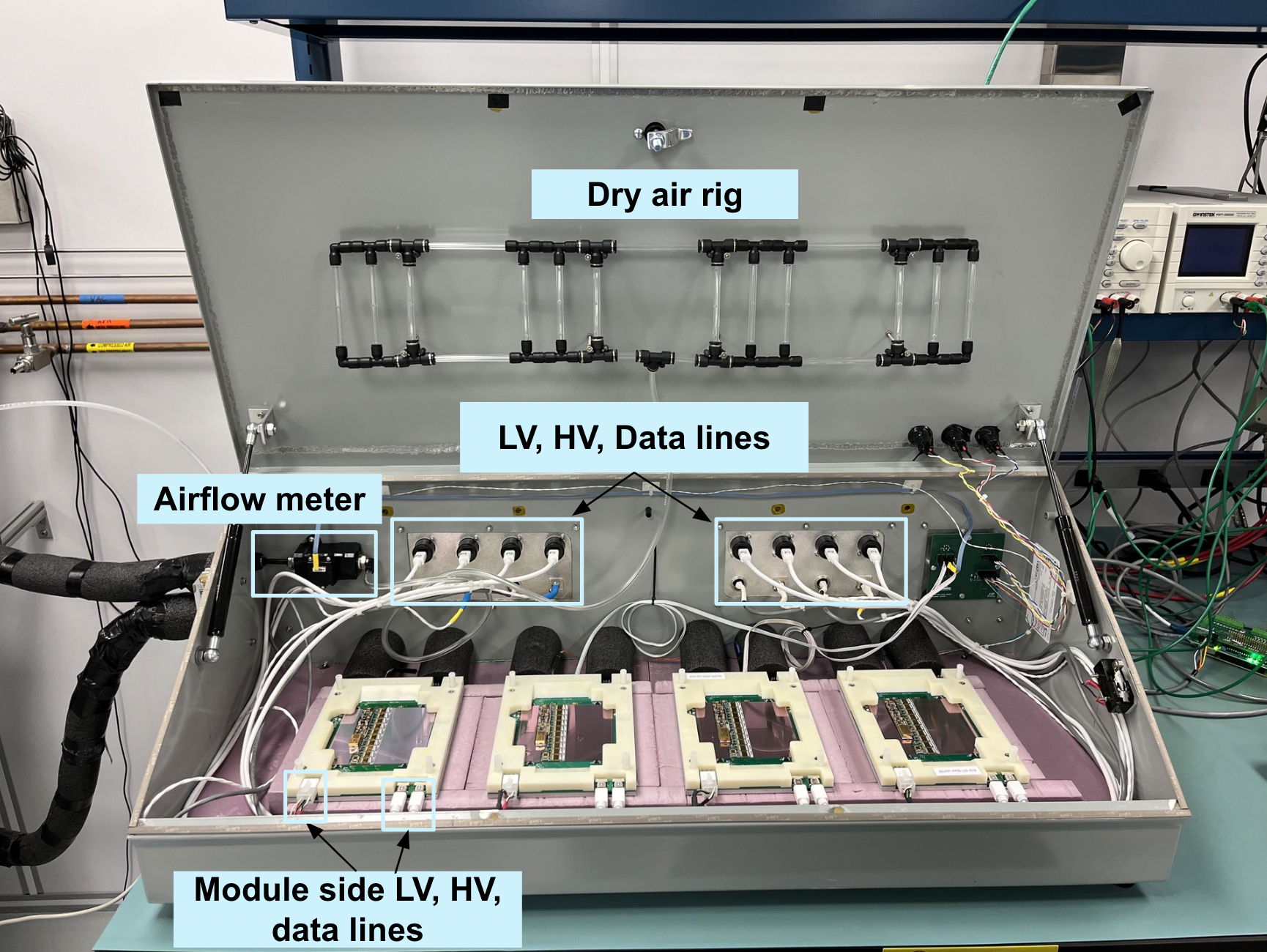}
\caption{Opened coldjig with four Long Strip barrel modules attached.\label{fig:ColdboxInside}}
\end{figure}

A photo of the interior of a coldjig with four Long Strip barrel modules inside is shown in figure~\ref{fig:ColdboxInside}. The dry air rig directs dry air from the external pipes onto the modules. An airflow meter is installed in order to monitor the rate of airflow in Standard Cubic Feet per Minute (SCFM), which must be above 1.5 SCFM at all times. LV and HV are provided to the module via a Molex connection to a test frame. Through the same test frame, data from the module is read out through the back of the box, where it is connected to the FMC-0514-DP with DisplayPort cables.

\section{Thermal cycling}
\label{sec:TC}

Throughout a module's lifetime in the ITk, it will experience a number of temperature changes due to planned or unplanned detector shutdowns. It is imperative that modules are mechanically robust, and able to take high quality data over the course of the ITk lifetime. The effect of temperature changes on modules is emulated with thermal cycling: One thermal cycle is defined as the chiller starting at the required cold temperature, a cold electrical test, the chiller warming to the pre-defined warm temperature, a warm electrical test, and the chiller returning to the required cold temperature to begin another thermal cycle. This is achieved using setups identical or similar to that described in section~\ref{sec:setup}. This procedure is shown in figure~\ref{fig:TC_QC_diagram}, where the original QC procedure included an upper limit for the thermal cycle of +40$^{\circ}$C. However, while developing this procedure during pre-production, the upper limit was reduced to +20$^{\circ}$C after it was found that powered modules at +40$^{\circ}$C approached the glass transition temperature of the glue securing electronics to the sensor.\label{fig:TC_QC_diagram} Additionally, a nominal thermal gradient of 2.5$\degree$C per minute is defined as part of the standard procedure. This is the final step of the module QC program: all modules must pass a post-cycling electrical test, and an I-V before being placed on a local support structure.

\begin{figure}[htbp]
\centering
\includegraphics[width=\textwidth]{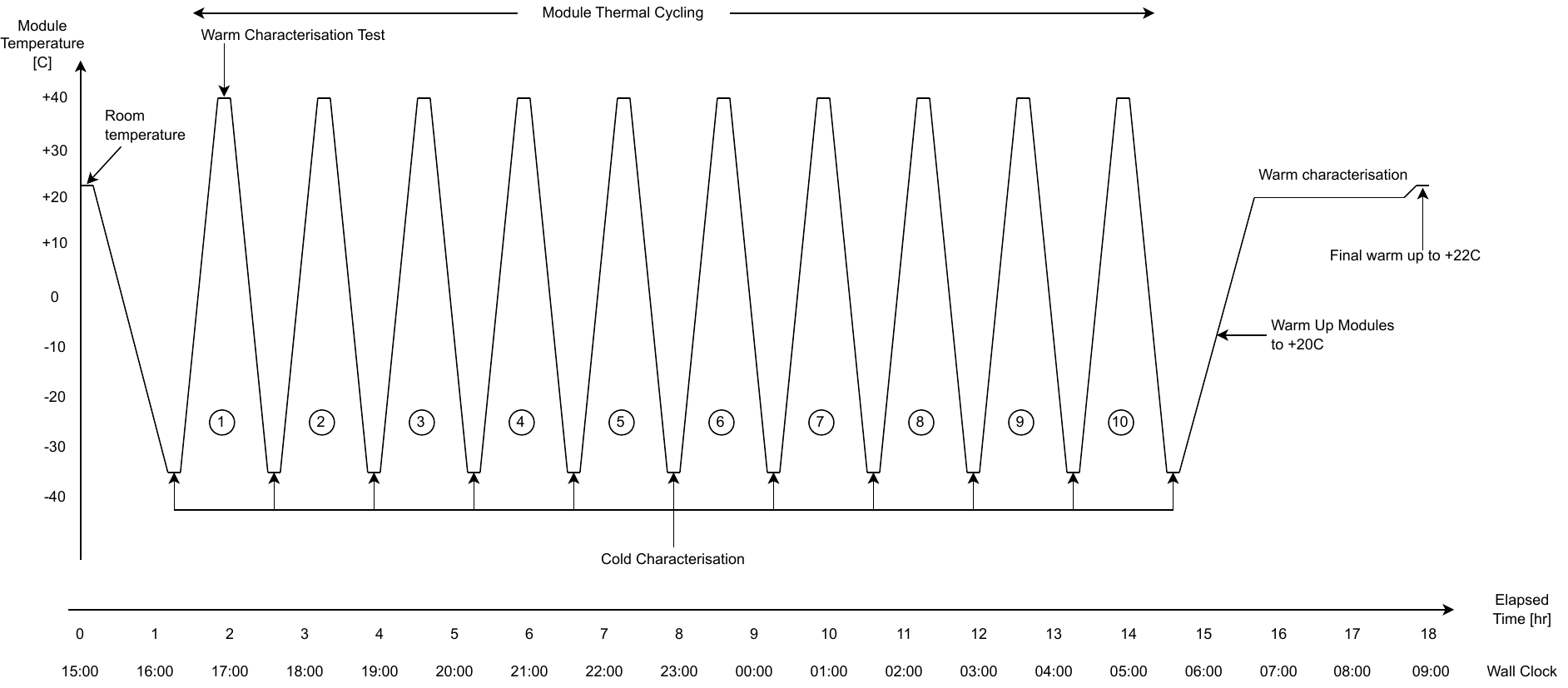}
\caption{The thermal cycling QC procedure. The diagram shows an upper limit for the thermal cycle of +40$^{\circ}$C, as per the original QC plan.}
\end{figure}

\subsection{DAQ tests}

As part of the module QC program, all modules must pass a set of electrical and mechanical tests, some of which were mentioned and included in the thermal cycling sequence defined in section~\ref{sec:TC}. I-V curves are taken to -550V at room temperature, and the module fails if it exhibits breakdown (a sudden spike in current) before reaching -500 V~\cite{IV_specs}. Electrical tests are also performed to characterize the module's noise and gain.

The ABCStar~\cite{ABCStar}, a binary readout chip used for strip readout, provides a charge injection circuit in order to characterise channels. By repeatedly injecting well-timed charges of equal amplitude and performing a scan of the threshold voltage in the front-end discriminator, a measure of the output noise in a channel is obtained. Several such scans with different injected charges gives a measure of the gain of a channel, which further allows relating the output noise to the equivalent charge in the input stage of the chip that would create similar output noise. The equivalent noise charge (ENC) will be expressed in units of $e$. Further details can be found in section 3.2.6 of \cite{ABC130}.    

A channel (corresponding to a single strip) is marked as bad if it has atypical gain, a noise measurement that is incompatible with the projected performance of the ITk after the expected irradiation, or is an outlier in gain or noise compared to the rest of the channels on the same chip. Similarly, a whole chip may fail the test if the average gain or noise on the chip is atypical, or if there are large variations between channels. 
A module fails the electrical test if one chip fails, if more then 2\% of its channels are marked as bad, or if more than eight consecutive channels are marked as bad. More details on module electrical test QC can be found in~\cite{ITkstripsQC}.



\subsection{Performance Trends}
\label{subsection:TC_trends}


The pre-production phase of the ITk Strips project has provided a chance to carefully examine the proposed thermal cycling procedure, where dedicated studies including the trends of output channel noise as a function of thermal cycle test number were performed. While DAQ quantities including noise and gain were monitored, leakage current was also constantly monitored in order to detect any unusual sensor behavior, or clear indications of breakdown.

\begin{figure}
    \centering
    \includegraphics[width=6in]{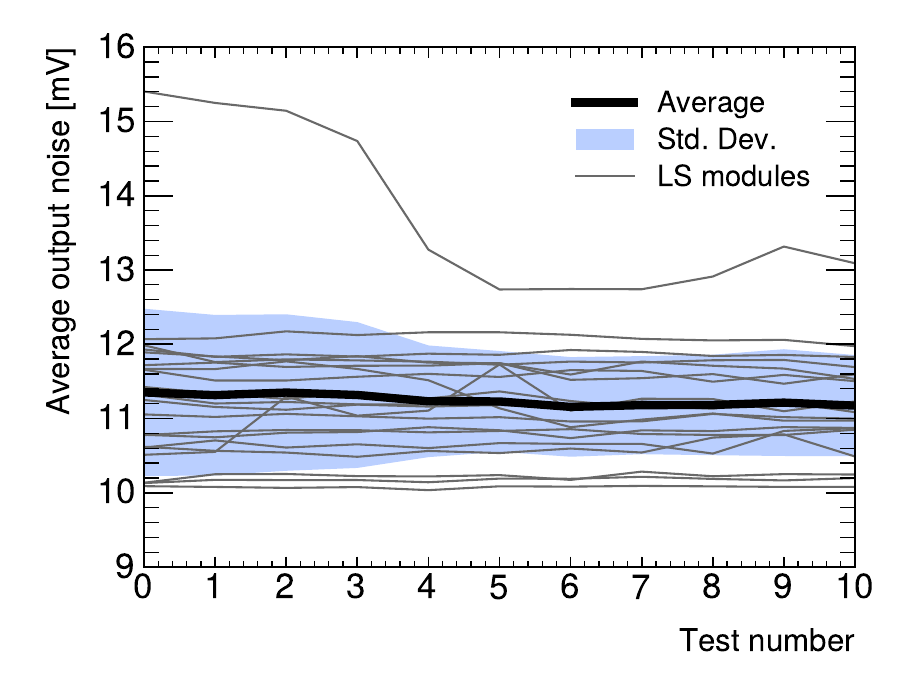}
    \caption{The average output noise of all strips in LS modules, during the 11 cold tests of thermal cycling (Test number 0 corresponds to the cold test before the first thermal cycle, see figure \ref{fig:TC_QC_diagram}). The thin grey lines show trends of 20 individual pre-production LS modules, along with the average and standard deviation between all modules shown by the black line and blue area respectively.}
    \label{fig:outputnoise-TC-trends}
\end{figure}

Results from these thermal cycling tests as a function of test number indicate whether there is any clear degradation in electrical performance as a consequence of the thermal stress. Figure \ref{fig:outputnoise-TC-trends} shows the development of output noise in 20 pre-production LS modules, at each of the 11 cold tests. Most individual modules demonstrate a stable noise that is almost unchanged in each of the cold tests, and the average results of all modules is consistent with there being no gradual noise increase. One module shows above average noise during cold tests, which improves as more cold tests are taken. In contrast, the noise from this module's warm tests are stable. It is not understood why this particular behavior is observed for this module.

\section{Extreme thermal cycling}
\label{sec:Extreme_Thermal_Cycling}

While each module must pass the thermal cycling QC procedure outlined in section~\ref{sec:TC}, it must be established that ten thermal cycles is not close to the breaking point of modules. If modules begin to fail after eleven cycles, this would not be caught by the thermal cycling QC procedure. To establish this, an extreme reliability test was performed in which modules were thermally cycled over 100 times and evaluated electrically and mechanically. Additionally, while warming follows the standard thermal gradient of about 2.5$\degree$C per minute, when cooling the chiller had an average rate of about 4$\degree$C per minute, adding an additional stress factor in testing these modules. The objective of this test is to make sure modules are reliably robust to a number of temperature changes much greater than what is expected during their lifetimes at the HL-LHC, and with more extreme conditions, as a safety factor to the nominal thermal cycling QC. As described in section~\ref{sec:introduction}, this is also motivated by power cuts which are expected to occur during operations, about 3-5 times per year.

For this test, four pre-production Long Strip barrel modules with the current production design were thermally cycled, henceforth referred to as Module 1, Module 2, Module 3, and Module 4. These modules are representative in assembly of the pre-production phase of barrel modules. For these tests, the chiller temperature alternated from -40$\degree$C to $+$40$\degree$C for cold and warm tests. It should be noted that a chiller temperature of -40$\degree$C leads to a chuck temperature of about -35$\degree$C, the planned operational temperature of the ITk after sufficient radiation damage has been accumulated. During cycling, the sensors were constantly biased by the applied high voltage, even during temperature changes\footnote{The thermal cycling QC procedure was later changed to only include sensor biasing during DAQ tests, not during temperature changes, as this will not be done during operations.}. It should be noted that the full set of thermal cycles was split into chunks, spread out over several weeks.

Within the first ten thermal cycles, the leakage current of Module 4 reached its upper current limit, and hence this module failed the nominal thermal cycling QC outlined in section~\ref{sec:TC}. This module would never have been part of the final detector, and therefore only results from the other three modules will be described. It should be noted that this module came from a batch of modules where many exhibited high leakage currents, the cause of which is suspected to be the seepage of glue onto the modules' guard rings~\cite{Cole_GlueOnGR}. Additionally, it should be noted that these modules exhibit noise greater than the target noise values before any thermal cycling. Performing this extreme thermal cycling test on modules that exceed the maximum target noise serves as a ``worst case" test. 

\subsection{Module 1}
\label{subsec:BNL_Module_1}

Module 1 was thermally cycled 101 times in the coldjig described in section~\ref{sec:setup}. During its first 62 thermal cycles, no change in behavior or performance was seen. During its 63rd thermal cycle the module's leakage current increased from its baseline value of about 200-500 nA to about 1.5~$\upmu$A, and then during its 64th thermal cycle to about 2.5~$\upmu$A. The cold DAQ test from the 64th cycle indicates a few very noisy strips on the sensor's edge. 

\begin{figure}[htbp]
\centering
\includegraphics[width=\textwidth]{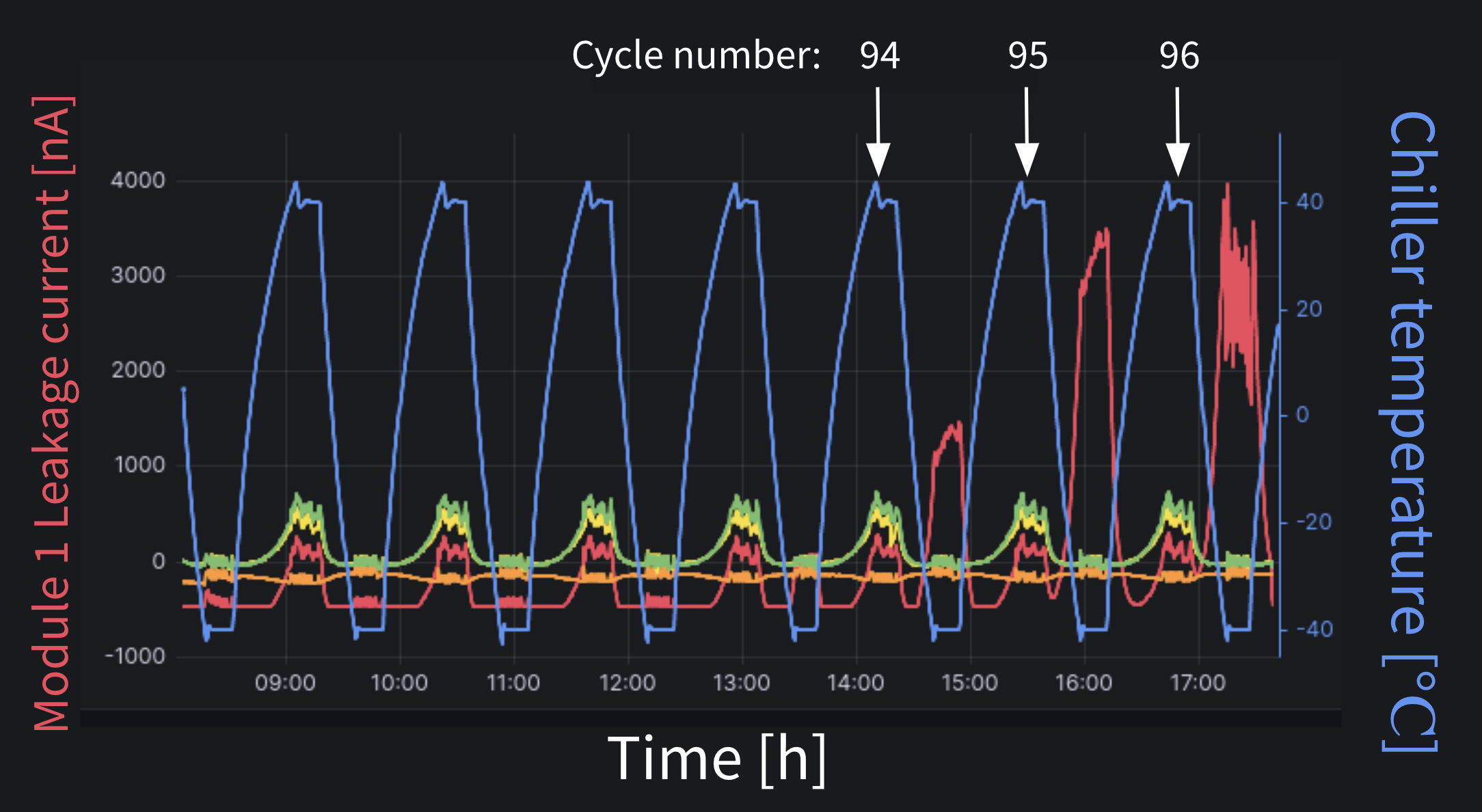}
\caption{Module 1 leakage current (in red) during thermal cycles 90 to 96. The leakage current from Module 2, Module 3, and Module 4 are shown in yellow, green, and orange respectively.\label{fig:BNL_Module_1_current}}
\end{figure}

During its next day of thermal cycling three days later, the first two thermal cycles did not show changes in behavior, but increased leakage current was again observed from cycles 66-70, this time up to about 4~$\upmu$A. The next time the module was cycled four days later, the module showed no changes in behaviour during cycles 71-80. On its second to last day of cycling, an instance of increasing leakage current while cooling down was observed during cycles 94-96 as shown in figure~\ref{fig:BNL_Module_1_current}, and relatively high leakage currents of about 4~$\upmu$A are seen during the final 5 cycles. In all of these instances, the current values are under the specification limit of 10~$\upmu$A. It should be noted that during HL-LHC operations, sensor current will increase by several orders of magnitude, and irradiation will often remove any breakdown. It is also noted that increasing sensor current during temperature changes while sensors are biased, similar to what was observed for Module 1, has been observed in other modules during pre-production. 

\begin{figure}[htbp]%
    \setcounter{subfigure}{0}
    \centering
    \subfloat[Stream 0: corresponds to strips under the hybrid.]{\includegraphics[clip, trim=0cm 0cm 0cm 0.1cm, width=0.485\textwidth]{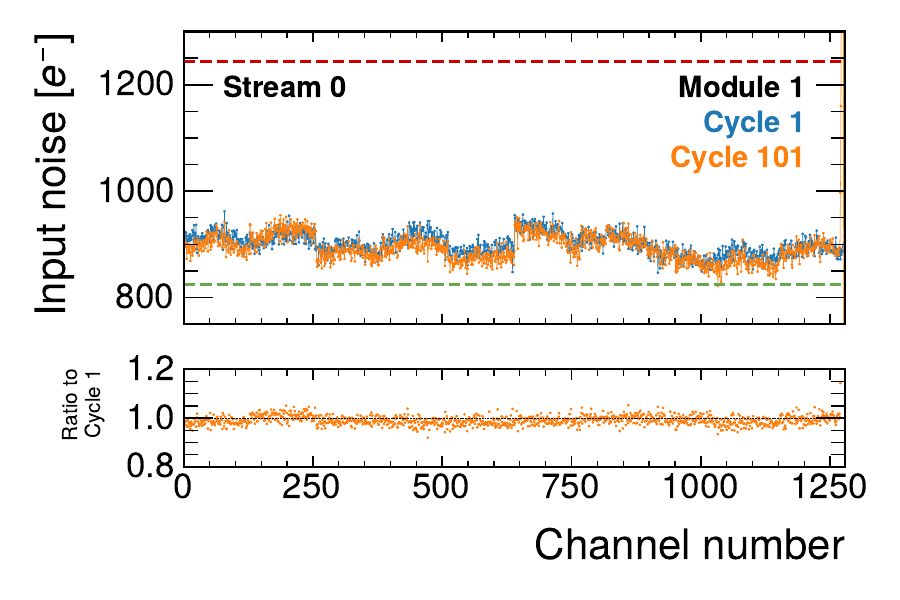}}%
    \hfill
    \subfloat[Stream 1: corresponds to strips away from the hybrid.]{\includegraphics[width=0.485\textwidth]{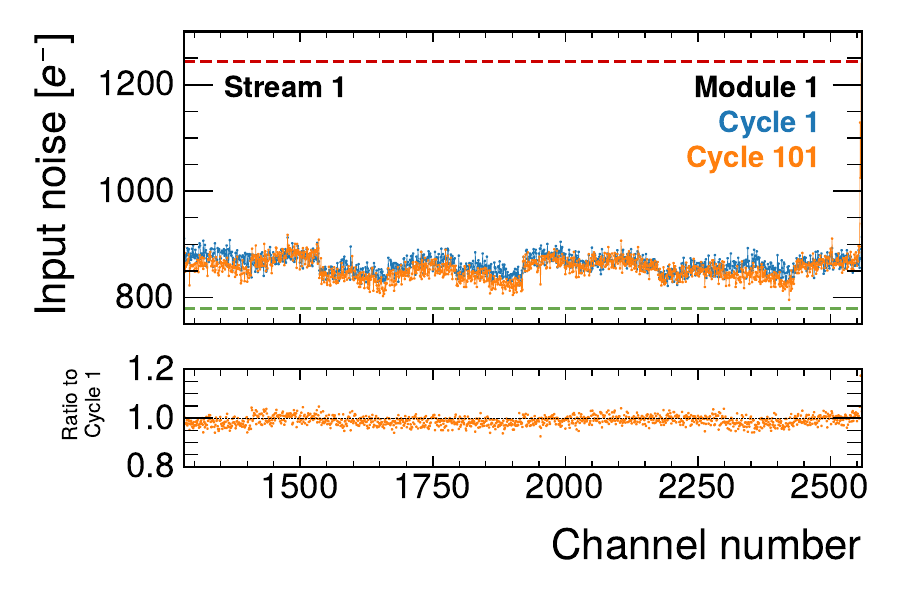}}%
    \caption{Comparison of Module 1 noise measured during first and final cold tests for streams 0 and 1.\label{fig:BNL_Module_1_Noise}}%
\end{figure}

A comparison of Module 1's noise during its first, and 101st cold tests is shown in figure~\ref{fig:BNL_Module_1_Noise} for both streams of strips. While there are a few noisy strips on the edges of each stream (near channel numbers 1280 and 2550), the vast majority of strips have the same measured noise within 5\% before and after cycling. Additionally, the measured noise is well below the maximum allowed noise of 1243 ENC. One would expect a higher noise measurement if taken at room temperature, but would not expect the noise to increase by the $\approx$ 300 ENC necessary to reach the threshold as noise increases of this magnitude are not typically seen in Long Strip modules between cold and room temperature tests. The measured noise exceeds the target room temperature noise in green, 824 (779) ENC for stream 0 (1), even when measured cold, indicating this module is noisier than expected. Upon inspection after testing, this module was found to have glue on the guard ring, which is known to increase sensor current measurement.

\begin{figure}[htbp]
\centering
\includegraphics[width=\textwidth]{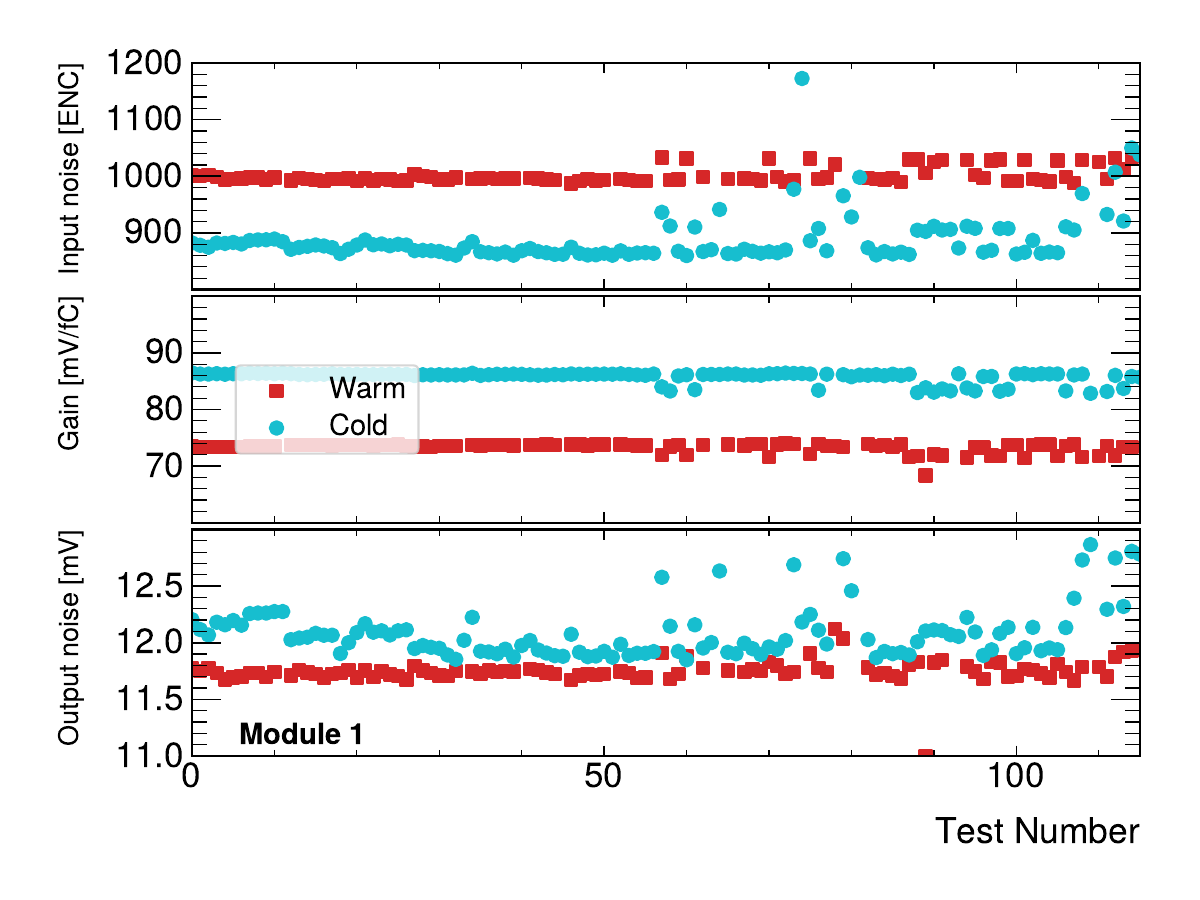}
\caption{Module 1 input noise, gain, and output noise as a function of test number, separated by warm (Chiller at +40$^{\circ}$C) and cold (Chiller at -40$^{\circ}$C) tests. Note that there were more tests taken than number of thermal cycles, as sometimes a day of cycling would end on a warm test without completing a cycle, or one day would end with a cold test and the next day would start with a cold test.\label{fig:BNL_Module_1_Trends}}
\end{figure}

Trends of Module 1's input noise, gain, and output noise during warm and cold tests as a function of test number are shown in figure~\ref{fig:BNL_Module_1_Trends}. These trends show a roughly steady increase in average input and output noise vs. test number, indicating that the module is relatively stable.

\begin{figure}[htbp]%
    \setcounter{subfigure}{0}
    \centering
    \subfloat[I-V curve.\label{fig:BNL_Module_1_IV}]{\includegraphics[width=0.485\textwidth]{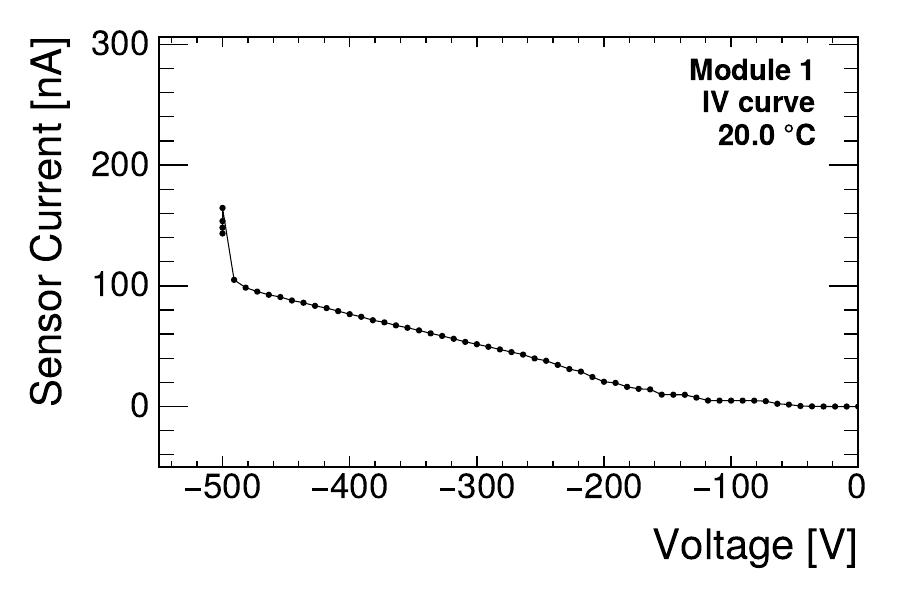}}%
    \hfill
    \subfloat[Measured height as a function of module position, considering the same orientation as shown in figure~\ref{fig:module_image}.\label{fig:BNL_Module_1_Bow}]{\includegraphics[width=0.485\textwidth]{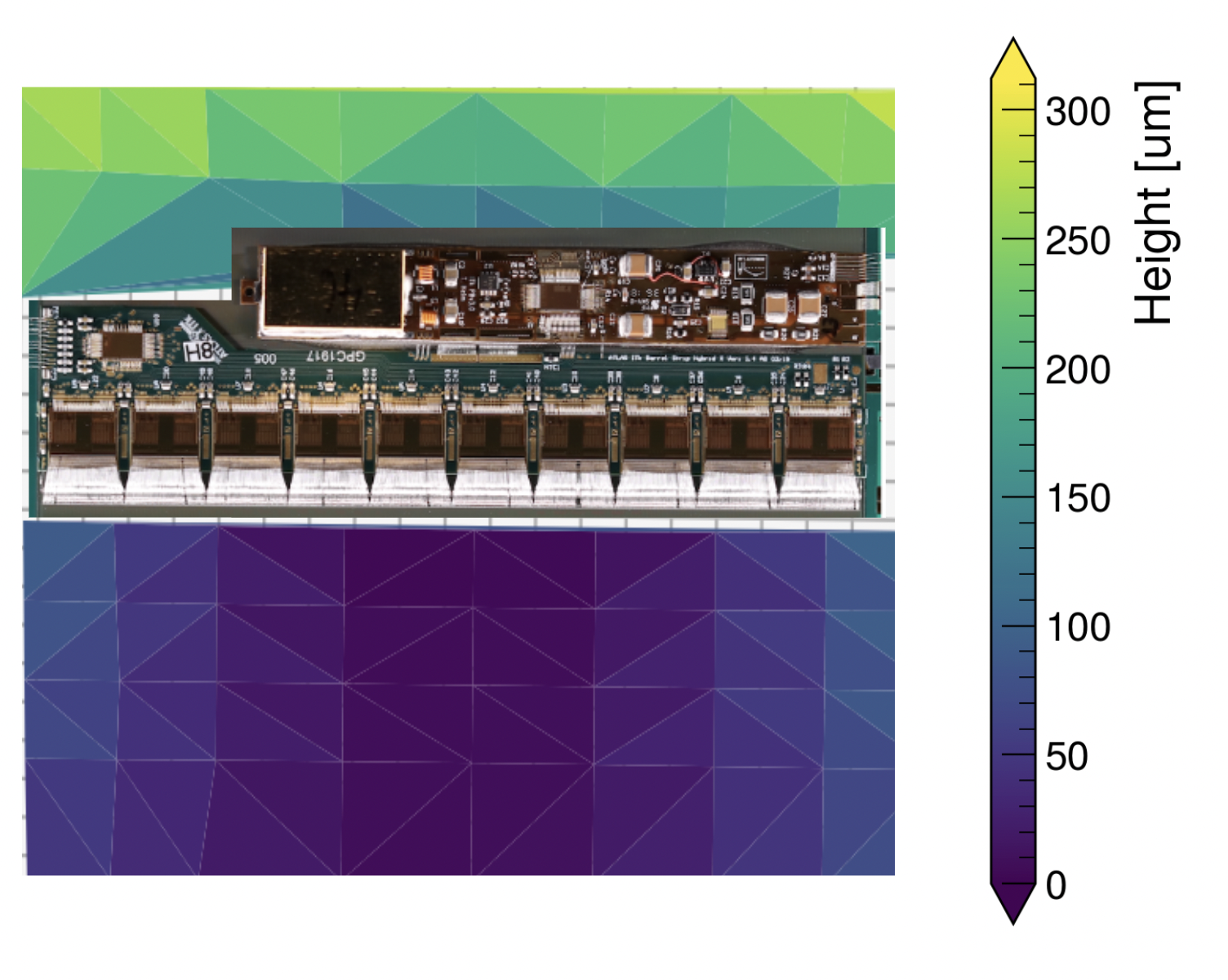}}%
    \caption{Post-cycling measurements of Module 1.}%
\end{figure}

To further evaluate electrical performance, an I-V curve was taken at room temperature after thermal cycling, and is shown in figure~\ref{fig:BNL_Module_1_IV}. Around -500V, the module begins to exhibit breakdown-like behavior, indicating it may have failed the nominal module QC which requires no breakdown before reaching -500V. However, this I-V shows signs of ``training", as the current around -500V goes down with time as seen during its final three measurements, indicating the observed signs of breakdown may not be permanent. Furthermore, this module exhibits a large operational voltage range before reaching -500V. During operation, a module is expected to experience a convolution of effects from thermal cycling, radiation, and the ATLAS magnetic field. However, if behavior like this is seen during operation, the magnitude of the operating voltage can be decreased to move away from the breakdown region. After testing, this module was discovered to have pinholes. This can potentially explain the irregular step-like shape of the I-V curve.

A post-cycling metrology measurement was also taken, and is shown in figure~\ref{fig:BNL_Module_1_Bow}. Deformations are seen in the module in both the horizontal and vertical planes, and a maximum height of about 300 microns was measured. This exceeds the acceptable metrology specifications, as the corners can be no greater than 150 microns above the center of the module, and no less than 50 microns below the center of the module. This measurement, along with comparisons to shape measurements of modules thermally cycled to warm temperatures of $+$40$\degree$C and $+$20$\degree$C led to the decision to change the thermal cycling warm temperature in the QC procedure to $+$20$\degree$C, where a much less severe shape distortion was observed.

\subsection{Module 2}
\label{subsec:SCIPP_Module_1}

\begin{figure}[htbp]
\centering
\includegraphics[width=\textwidth]{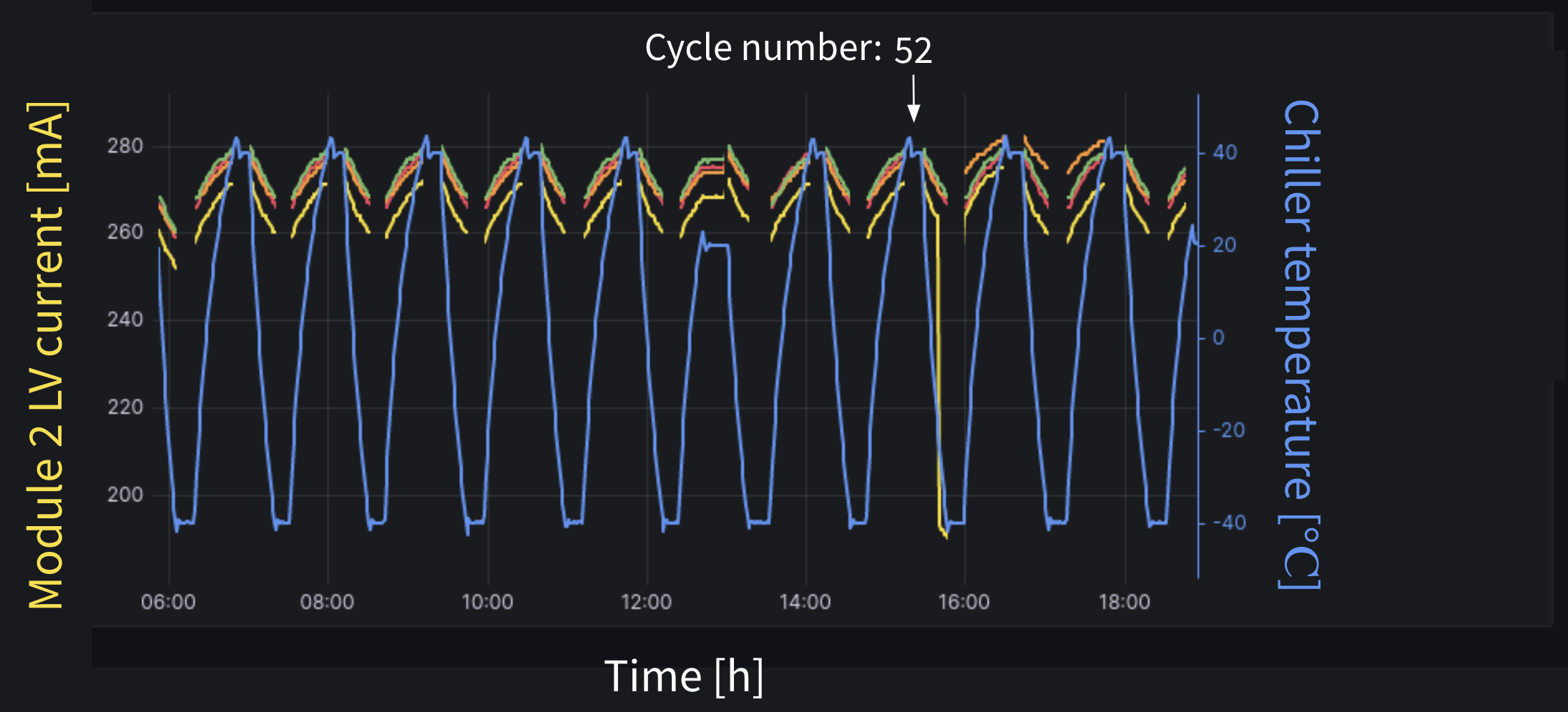}
\caption{Module 2 LV current from cycles 45 to 54 (in yellow). The LV current for Module 1, Module 3, and Module 4 are shown in red, green, and orange, and stay around the expected value of 280 mA, with changes of about 10 mA due to cooling and warming. \label{fig:SCIPP_Module_1_LV_current}}
\end{figure}

Module 2 was thermally cycled 101 times\footnote{Before this, this module was thermally cycled 4 additional times at a higher bias voltage.} in the coldjig described in section~\ref{sec:setup}. For the first 51 cycles of this module, no significant deviations in sensor noise or current were observed. Starting with the 52nd cycle, intermittent instability was observed in input noise and LV current draw. An example of this is shown in figure~\ref{fig:SCIPP_Module_1_LV_current}. This suggests that either the pedestal trim was not properly applied during these tests, or the ASICs were not properly powered.


\begin{figure}[htbp]%
    \setcounter{subfigure}{0}
    \centering
    \subfloat[Stream 0: corresponds to strips under the hybrid.]{\includegraphics[width=0.485\textwidth]{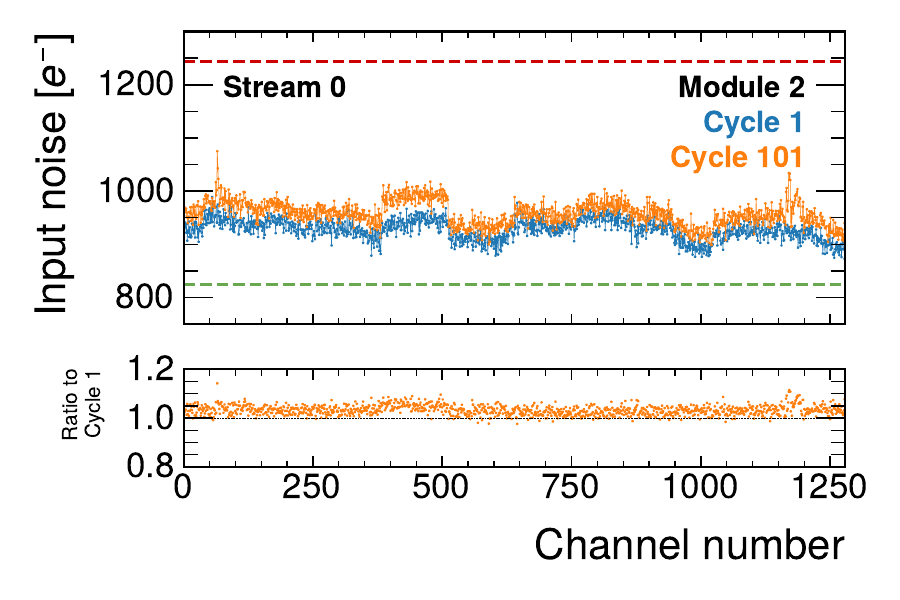}}%
    \hfill
    \subfloat[Stream 1: corresponds to strips away from the hybrid.]{\includegraphics[width=0.485\textwidth]{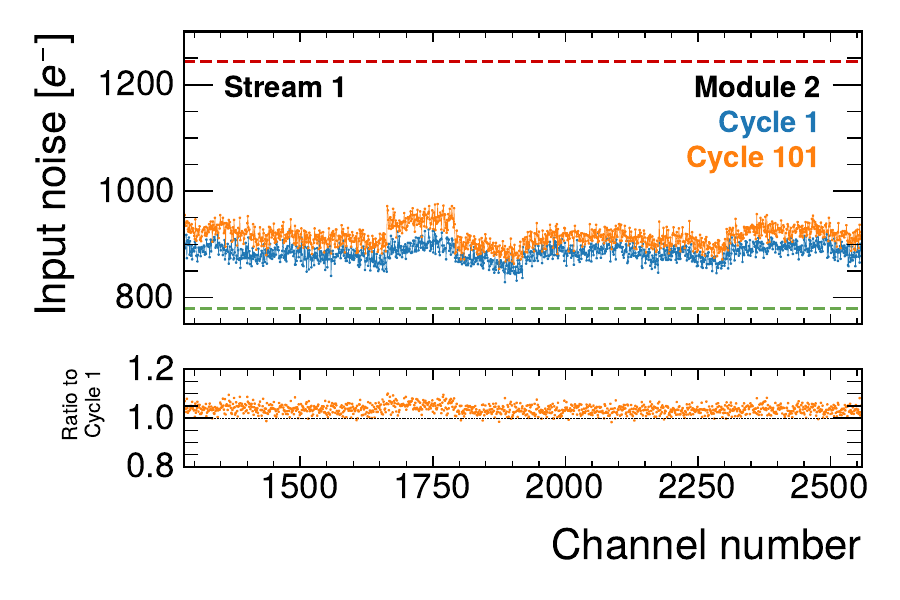}}%
    \caption{Comparison of Module 2 noise measured during first and final cold tests for streams 0 and 1. \label{fig:SCIPP_Module_1_Noise}}%
\end{figure}

A comparison of input noise between its first and final cold test is shown in figure~\ref{fig:SCIPP_Module_1_Noise}. While there are a few relatively noisy strips around channels 60 and 1175 (also seen in the output noise results), and there is a roughly constant increase in noise among all strips of this module after thermal cycling, the majority of strips are less than 5\% noisier after cycling. As with Module 1, the measured noise is well below the maximum allowed room temperature noise of 1243 ENC. Also similar to Module 1, the measured noise exceeds the target room temperature noise in green, 824 (779) ENC for stream 0 (1), even when measured cold, indicating this module is noisier than expected.

\begin{figure}[htbp]
\centering
\includegraphics[width=\textwidth]{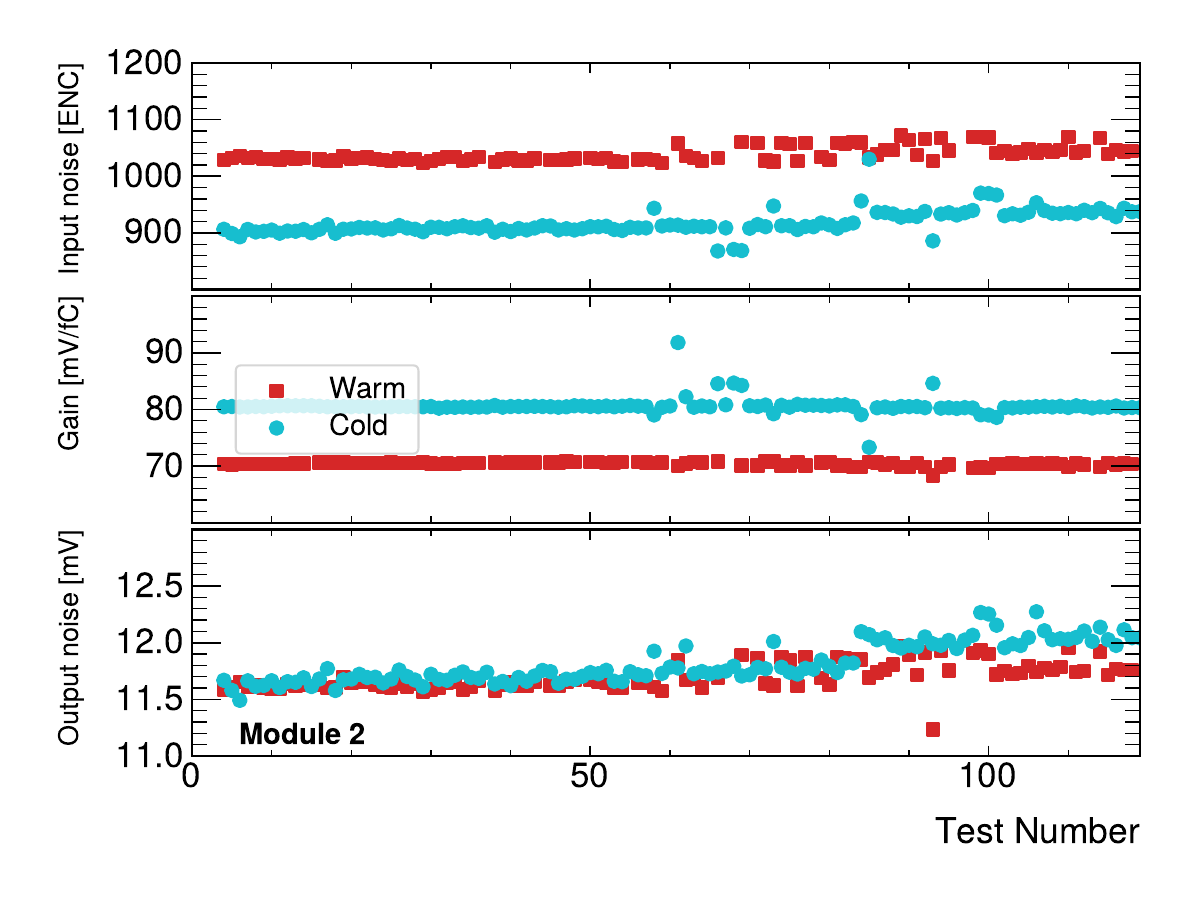}
\caption{Module 2 input noise, gain, and output noise as a function of test number, separated by warm (Chiller at +40$^{\circ}$C) and cold (Chiller at -40$^{\circ}$C) tests. Note that there were more tests taken than number of thermal cycles, as sometimes a day of cycling would end on a warm test without completing a cycle, or one day would end with a cold test and the next day would start with a cold test.\label{fig:SCIPP_Module_1_Trends}}
\end{figure}

\begin{figure}[htbp]%
    \setcounter{subfigure}{0}
    \centering
    \subfloat[I-V curve.\label{fig:SCIPP_Module_1_IV}]{\includegraphics[width=0.485\textwidth]{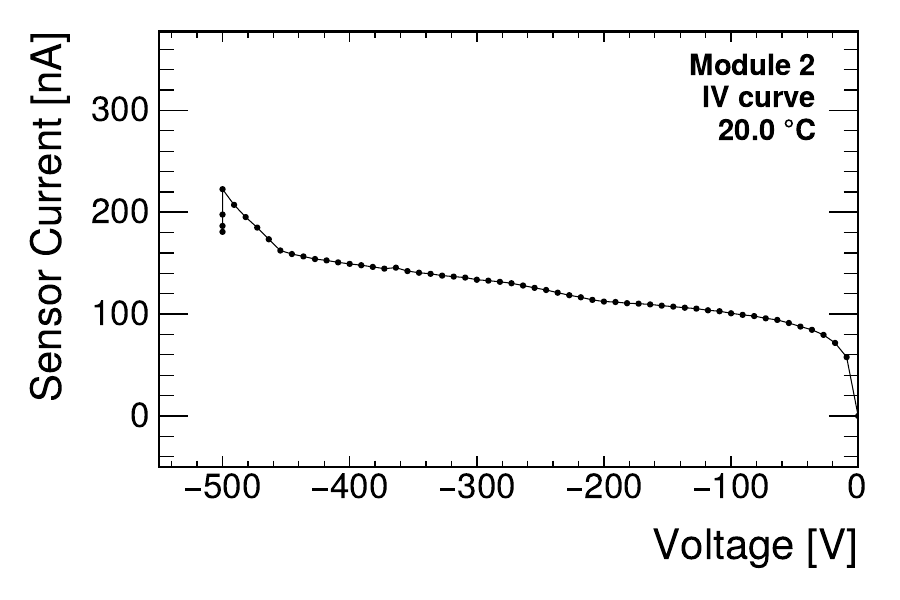}}%
    \hfill
    \subfloat[Measured height as a function of module position, considering the same orientation as shown in figure~\ref{fig:module_image}.\label{fig:SCIPP_Module_1_Bow}]{\includegraphics[width=0.485\textwidth]{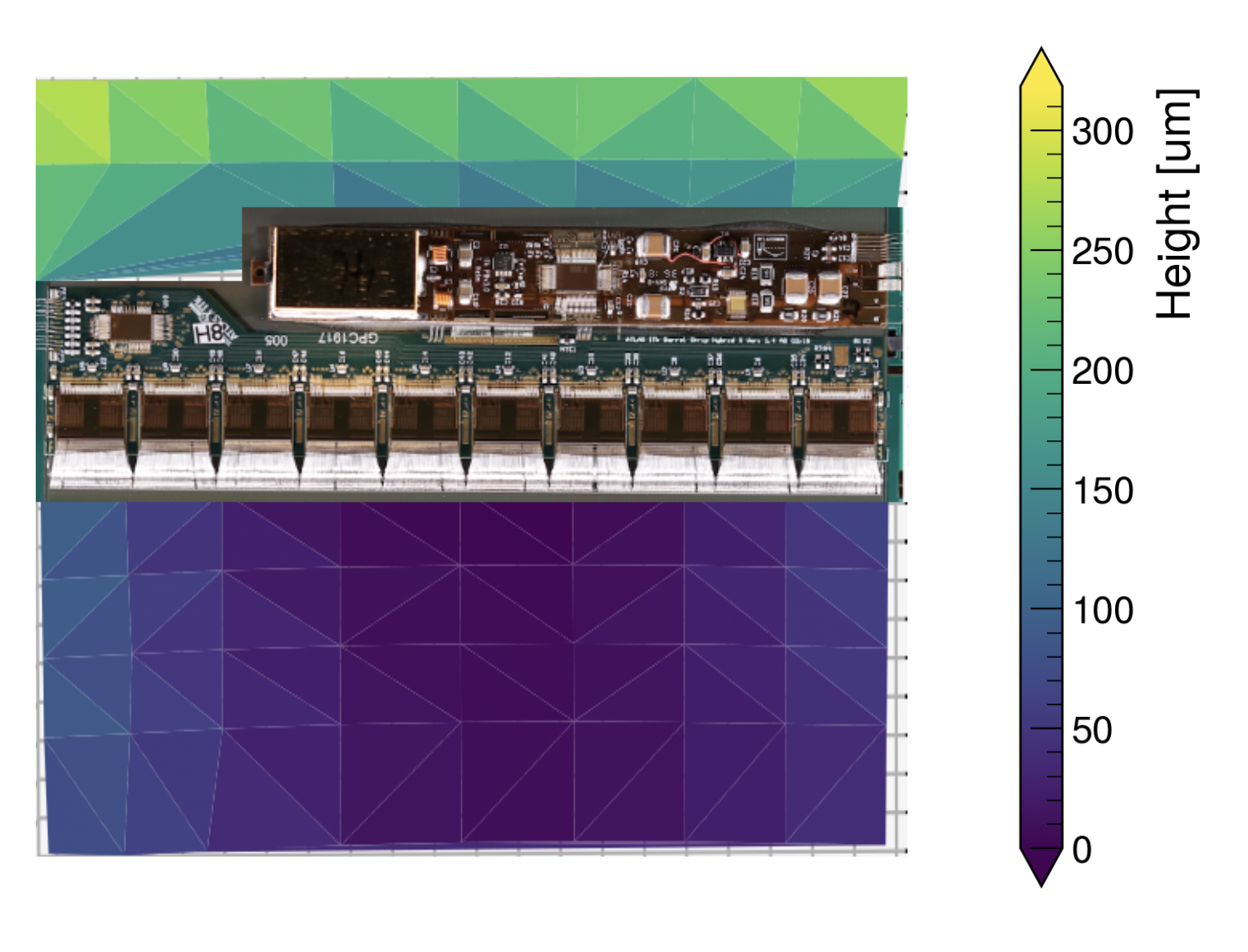}}%
    \caption{Post-cycling measurements of Module 2.}%
\end{figure}

Trends of Module 2's input noise, gain, and output noise during warm and cold tests as a function of test number are shown in figure~\ref{fig:SCIPP_Module_1_Trends}. These trends show a slow rise in average input and output noise vs. test number. While this is non-desirable behavior, noise is still well within specification and 100 cycles is far beyond the expected detector induced stress. Finally, an I-V curve was taken after cycling, and is shown in figure~\ref{fig:SCIPP_Module_1_IV}.

Similar to Module 1, this module begins to show increased leakage current as it approaches -500V. It is possible that this module would have failed the nominal QC, as it exhibits breakdown-like behavior before -500V. However, as was the case with Module 1, this I-V shows signs of training indicating its observed signs of breakdown may not be permanent. Also similar to Module 1, it may still be possible to operate this module depending on how the rest of the in-situ effects alter the module's electrical performance.

A post-cycling metrology measurement was also taken, and is shown in figure~\ref{fig:SCIPP_Module_1_Bow}. A similar shape measurement to that of Module 1 is seen. 

\subsection{Module 3}
\label{subsec:SCIPP_Module_2}

Module 3 was thermally cycled 101 times\footnote{Before this, this module was thermally cycled 4 additional times at a higher bias voltage.} in the coldjig described in section~\ref{sec:setup}. This module did not exhibit any significant changes during thermal cycling. 

\begin{figure}[htbp]%
    \setcounter{subfigure}{0}
    \centering
    \subfloat[Stream 0: corresponds to strips under the hybrid.]{\includegraphics[width=0.485\textwidth]{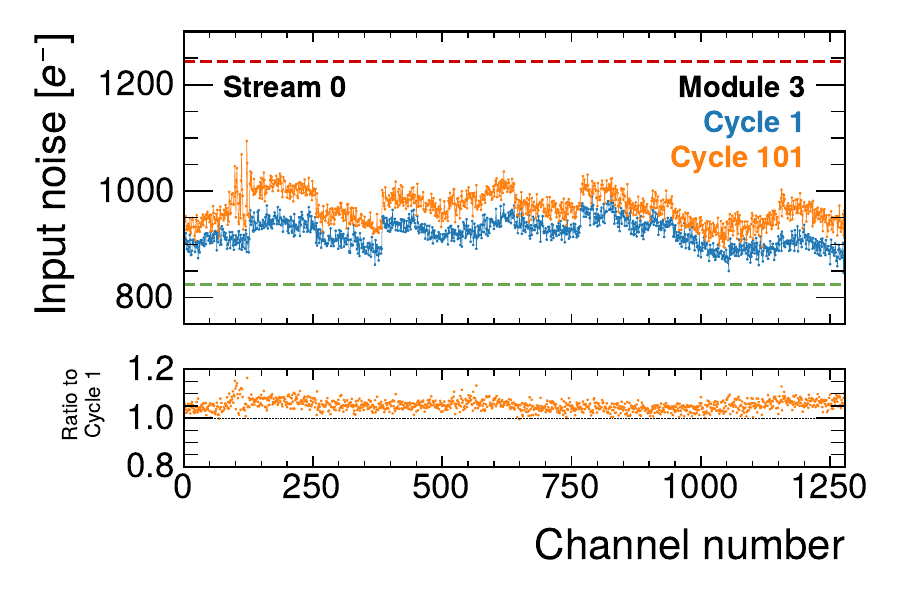}}%
    \hfill
    \subfloat[Stream 1: corresponds to strips away from the hybrid.]{\includegraphics[width=0.485\textwidth]{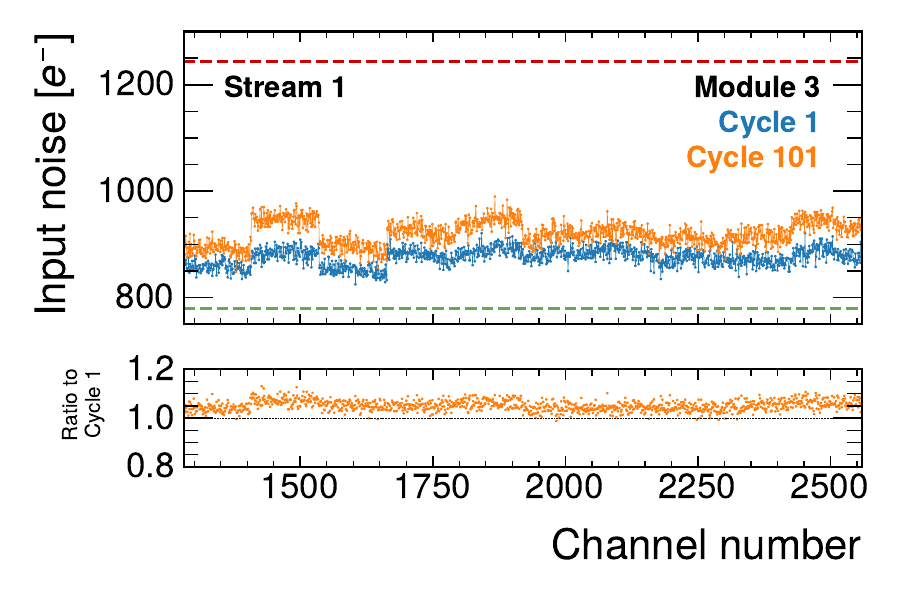}}%
    \caption{Comparison of Module 3 noise measured during first and final cold tests for streams 0 and 1. \label{fig:SCIPP_Module_2_Noise}}%
\end{figure}  

A comparison of input noise between its first and final cold test is shown in figure~\ref{fig:SCIPP_Module_2_Noise}. Similar behavior is seen to that of Module 2: there are a few relatively noisy areas around channels 100 and 1175, and a roughly constant increase of about 5\% in noise among strips after thermal cycling. As with Module 1 and Module 2, the measured noise is well below the maximum allowed room temperature noise of 1243 ENC. Additionally, the measured noise exceeds the target room temperature noise in green, 824 (779) ENC for stream 0 (1), even when measured cold, indicating this module was noisier than expected before any thermal cycling.

\begin{figure}[htbp]
\centering
\includegraphics[width=\textwidth]{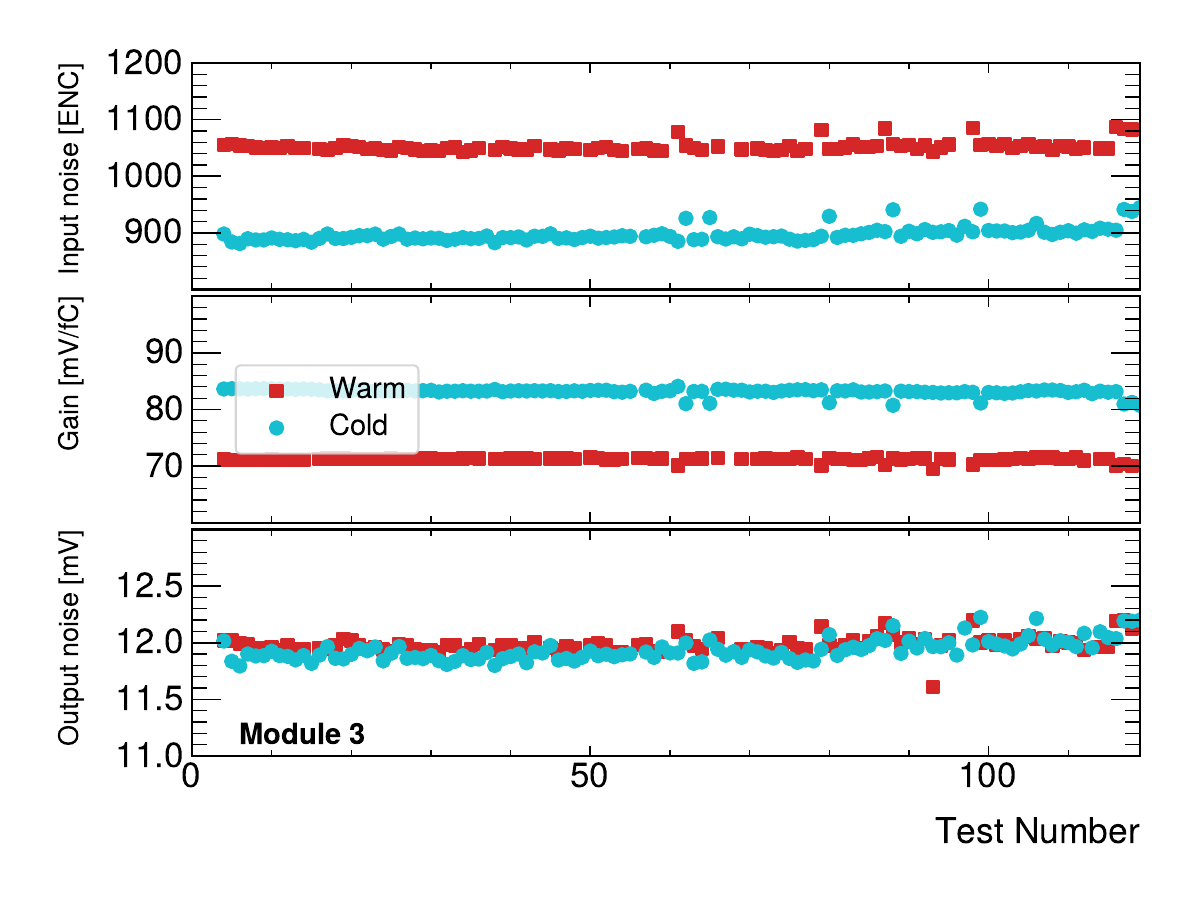}
\caption{Module 3 input noise, gain, and output noise as a function of test number, separated by warm (Chiller at +40$^{\circ}$C) and cold (Chiller at -40$^{\circ}$C) tests. Note that there were more tests taken than number of thermal cycles, as sometimes a day of cycling would end on a warm test without completing a cycle, or one day would end with a cold test and the next day would start with a cold test.\label{fig:SCIPP_Module_2_InputNoiseTrend}}
\end{figure}

Trends of this module's input noise, gain, and output noise during warm and cold tests as a function of test number are in figure~\ref{fig:SCIPP_Module_2_InputNoiseTrend}. Noise is consistently higher on average for warm tests, as expected. Additionally, these trends show a roughly steady increase in average input noise vs. test number, indicating that the module is relatively stable, similar to Module 2.

\begin{figure}[htbp]%
    \setcounter{subfigure}{0}
    \centering
    \subfloat[I-V curve.\label{fig:SCIPP_Module_2_IV}]{\includegraphics[width=0.485\textwidth]{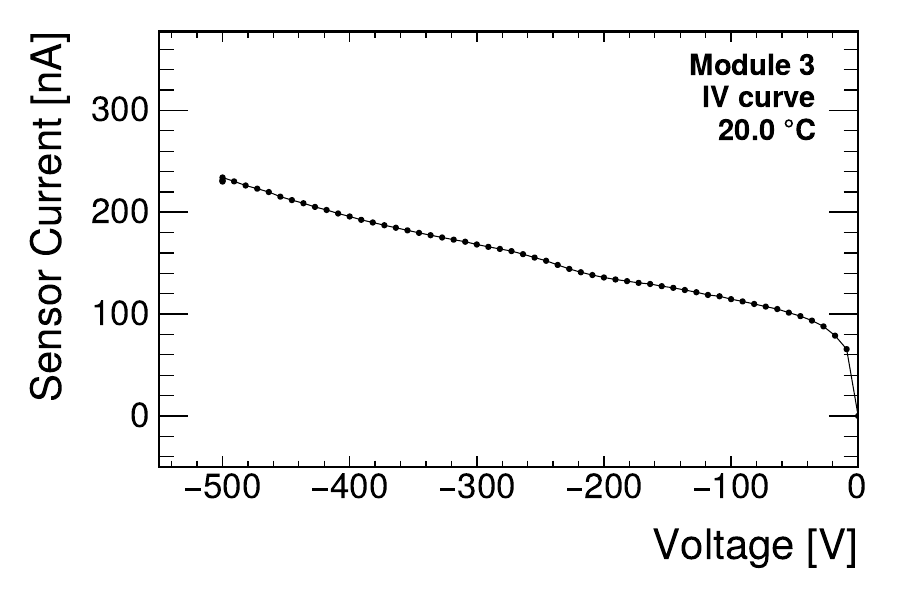}}%
    \hfill
    \subfloat[Measured height as a function of module position, considering the same orientation as shown in figure~\ref{fig:module_image}.\label{fig:SCIPP_Module_2_Bow}]{\includegraphics[width=0.485\textwidth]{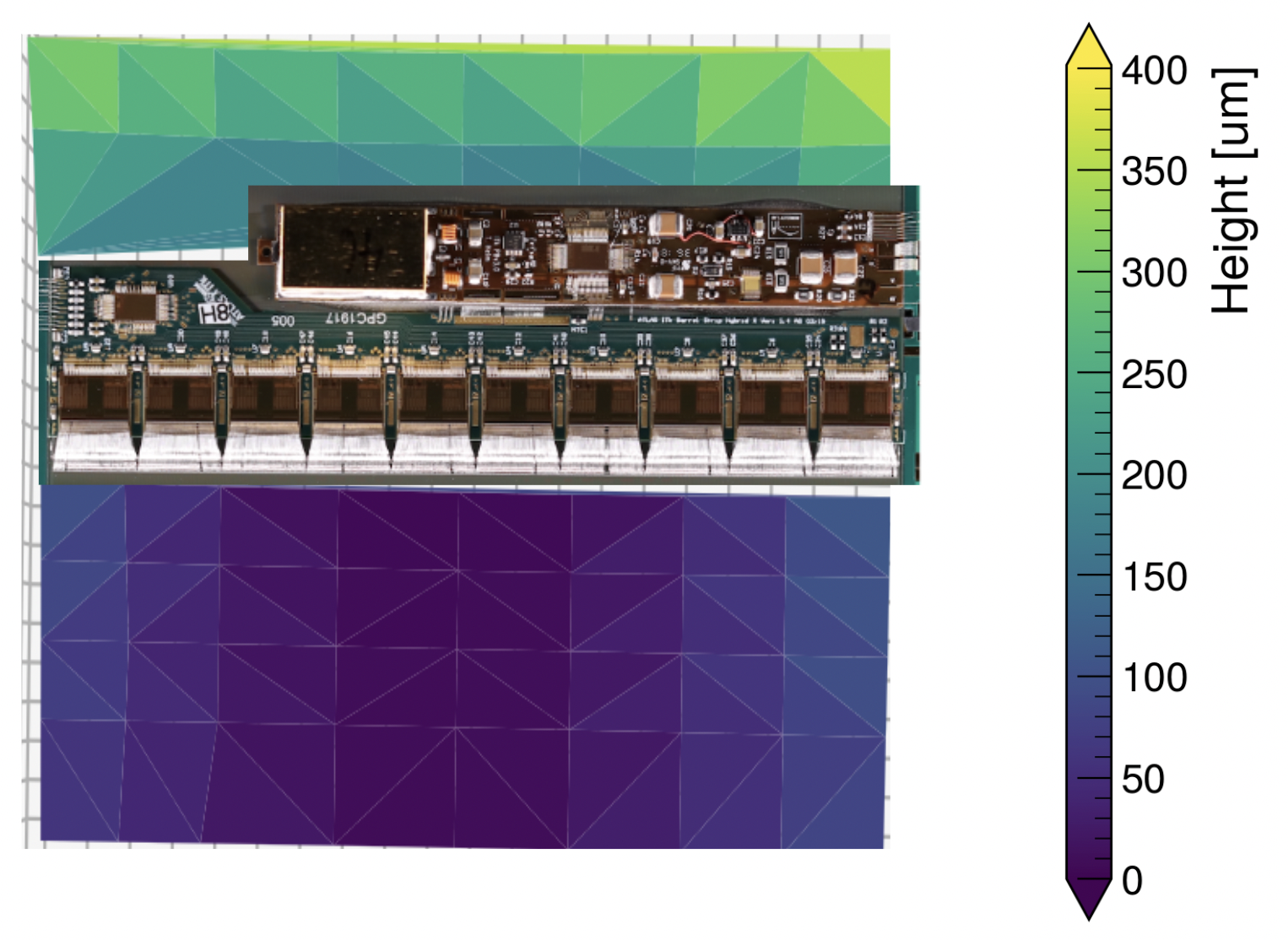}}%
    \caption{Post-cycling measurements of Module 3.}%
\end{figure}

Finally, an I-V curve was taken after cycling, and is shown in figure~\ref{fig:SCIPP_Module_2_IV}. This module does not show any indications of breakdown, and would have passed the I-V step of the nominal module QC. 

A post-cycling metrology measurement was also taken, and is shown in figure~\ref{fig:SCIPP_Module_1_Bow}. A similar shape measurement to that of Module 1 and Module 2 is seen. 

\section{Conclusions}
\label{sec:conclusions}

As part of the ITk Strips module QC procedure, all modules must be thermally cycled 10 times and pass a set of noise, gain, IV, and mechanical tests before being placed onto a local support structure. The barrel pre-production phase has allowed for a chance to carefully study module behavior under these test, and has shown many modules to have stable behavior.

In order to ensure modules are expected to remain operational with a safety factor on the number of temperature changes, partially motivated by expected power cuts during operations, a thermal cycling reliability test was performed on 4 Long Strip barrel modules, in which they were thermally cycled 100 times. One module failed within the first 10 cycles, and therefore would have failed the nominal QC and would not have been installed onto a local support structure. For Modules 2 and 3, a noise increase on order of about 5\% is seen after 100 thermal cycles, but this remains within specifications. Modules 1 and 2 exhibit breakdown-like behavior just before reaching -500V after 100 cycles, but this is not considered problematic as the modules still have large operational ranges, and we do not expect the detector to see this many cycles. This serves as a first test towards understanding the long term reliability of ITk Strips barrel modules to thermal cycling. Future plans include performing this test on more modules to gather more statistics.




\begin{thebibliography}{99}

\bibitem{HLLHC_TDR}
O. Aberle et al., \emph{High-Luminosity Large Hadron Collider (HL-LHC): Technical design report}, CERN
Yellow Reports: Monographs, \href{https://cds.cern.ch/record/2749422}{CERN-2020-010} (2020).

\bibitem{ITk_Layout}
The ATLAS Collaboration, \emph{Expected tracking and related performance with the updated ATLAS Inner Tracker layout at the High-Luminosity LHC}, \href{https://cds.cern.ch/record/2776651/}{ATL-PHYS-PUB-2021-024}.

\bibitem{Orr:2023eb}
R. Orr et al., \emph{Experimental Study and Empirical Modeling of Long Term Annealing of the ATLAS18 Strip Sensors}, \href{https://pos.sissa.it/420/043/}{https://pos.sissa.it/420/043/}

\bibitem{ABC130}
L. Poley et al., \emph{The ABC130 barrel module prototyping
programme for the ATLAS strip tracker}, \href{https://iopscience.iop.org/article/10.1088/1748-0221/15/09/P09004}{2020 JINST 15 P09004} (2020)

\bibitem{TDR}
ATLAS collaboration, \emph{Technical Design Report for the ATLAS Inner Tracker Strip Detector}, \href{https://cds.cern.ch/record/2257755}{CERN-LHCC-2017-005} (2017)

\bibitem{ModuleImage}
K. Affolder et al., \emph{Automated visual inspection and defect detection of large-scale silicon strip sensors}, \href{https://iopscience.iop.org/article/10.1088/1748-0221/17/03/P03026}{2022 JINST 17 P03026} (2022)

\bibitem{NEXYS}
xilinx, \emph{Nexys Video FPGA: Trainer Board for Multimedia Applicaitons}, \href{https://www.xilinx.com/products/boards-and-kits/1-cfdwic.html}{https://www.xilinx.com/products/boards-and-kits/1-cfdwic.html} (2023)

\bibitem{FMC}
K.J.R. Cormier et al., \emph{Development of the front end amplifier circuit for the ATLAS ITk silicon strip detector}, 2021, JINST 16 P07061 (2021).

\bibitem{RaspPi}
Raspberry Pi, \emph{Buy a Raspberry Pi 4 Model B}, \href{https://www.raspberrypi.com/products/raspberry-pi-4-model-b/}{https://www.raspberrypi.com/products/raspberry-pi-4-model-b/} (2023)

\bibitem{IV_specs}
D. Rousso, et al., "Test and extraction methods for the QC parameters of silicon strip sensors for ATLAS upgrade tracker", NIM A 1045 (2023) 1676

\bibitem{ABCStar}
M.J. Basso et al., \emph{A Starry Byte — proton beam measurements of single event upsets and other radiation effects in ABCStar ASIC Versions 0 and 1 for the ITk strip tracker}, \href{https://iopscience.iop.org/article/10.1088/1748-0221/17/03/P03017/meta}{2022 JINST 17 P03017} (2022)

\bibitem{ITkstripsQC}
A. Tishelman-Charny on behalf of the ATLAS ITk collaboration, \emph{The quality control programme for ITk strip tracker module assembly}, \href{https://arxiv.org/abs/2401.17054}{2401.17054} (2024)

\bibitem{Cole_GlueOnGR}
C. Helling et al., \emph{Study of n-on-p sensors breakdown in presence of dielectrics placed on top surface},
\href{https://www.sciencedirect.com/science/article/pii/S0168900218310829?via%3Dihub}{Nucl. Instrum. Meth. A 924 (2019) 147} 

\end{thebibliography}
\end{document}